\documentclass[twoside,onecolumn,nofootinbib,showkeys]{revtex4} 
\usepackage{dcolumn}
\usepackage{bm}
\usepackage[utf8]{inputenc}
\usepackage{pgfplots}
\usepackage{amsmath}
\usepackage{dcolumn} 

\newcommand{\BibTeX}{B\kern-0.1emi\kern-0.017emb\kern-0.15em\TeX}
\newcommand{\XYpic}{$\mathrm{X\kern-0.3em\raisebox{-0.18em}{Y}}$-$\mathrm{pic}\,$}

\newcommand{\cl}{C \kern -0.1em \ell}  

\newcommand{\ed}{\end{document}}
\begin{document}

%
%
%
%
%
%
%
%
%
\title{Representations of Extended Carroll Group }
\author{G. X. A. Petronilo}
%
\address{
International Center of
Physics, Instituto de F\'isica, Universidade de Bras\'ilia,\\
70910-900, Bras\'ilia, DF, Brazil}
\email{gustavo.petronilo@aluno.unb.br}
%
\author{S.C. Ulhoa}
\address{%
International Center of
Physics, Instituto de F\'isica, Universidade de Bras\'ilia,\\
70910-900, Bras\'ilia, DF, Brazil.\\
Canadian Quantum Research Center,\\ 204-3002 32 Ave Vernon, BC V1T 2L7  Canada} 
\email{sc.ulhoa@gmail.com}
\author{A. E. Santana}
%
\address{
International Center of
Physics, Instituto de F\'isica, Universidade de Bras\'ilia,\\
70910-900, Bras\'ilia, DF, Brazil}

%
%
\date{\today}
\begin{abstract}
Carroll's group is presented as a group of transformations in a 5-dimensional space ($\mathcal{C}$) obtained by embedding the Euclidean space into a (4; 1)-de Sitter space. Three of the five dimensions of $\mathcal{C}$ are related to $\mathcal{R}^3$, and the other two to mass and time. A covariant formulation of Caroll's group, analogous as introduced by Takahashi to Galilei's group, is deduced. Unit representations are studied.
\vskip 4pt
\end{abstract}
\label{page:firstblob}

 \begin{abstract}
Carroll's group is presented as a group of transformations in a 5-dimensional space ($\mathcal{C}$) obtained by embedding the Euclidean space into a (4,1)-de Sitter space. Three of the five dimensions of $\mathcal{C}$ are related to $\mathcal{R}^3$, and the other two to mass and time. A covariant formulation of Caroll's group, analogous as introduced by Takahashi to Galilei's group, is deduced. Unit representations are studied.
\vskip 4pt
\end{abstract}
\label{page:firstblob}
\maketitle

 \section{Introduction}

It is well known that low speed physics is obtained as a limit of Poincar\'e transformations.  On the other hand, there is a symmetry distinct from the Galilean that can also be obtained from the latter, the so-called Carroll transformations. In fact such a limit was first described by Levy-Leblond in 1965 \cite{levy1965}. It was shown that the Carroll group is a subgroup of de Sitter group\cite{bacry1968}. Duval \emph{et al}, in 2014, constructed a Carroll space from a (d+1,1) de Sitter space analogous with the Galilean one, using Bargmann structures~\cite{duval2014,duval2017}. Since then the Carroll group has been getting a lot of attention mainly in the context of strings~\cite{duval2014,jessen2015,tolkachev2016,ciambelli2018,calo2018,gomis2020}. The difference between Galilei and Carroll groups is that the last one swaps the transformation of the $x$ and $t$ axes when compared to the first one. Indeed, Carroll transformations are given by
$$
\begin{aligned}[c]
      &\overline{x}^i = R^i_jx^j\\
       &\overline{t}= t-(R^i_jx^j)v_i\,.
\end{aligned}
$$
It is worth noting that while the space is absolute, time is relative. It is equivalent to relativistic approximation when space-like intervals are much greater that of time-like ones..In this paper we will construct a five-dimensional de Sitter space and show that the Carroll space is defined as a light cone coordinate of de Sitter one. It is interesting to note that the search for a covariant structure in non-relativistic physics has a long history.  In fact, since the advent of general relativity, a covariant formulation for Newtonian physics has been sought.  In this sense, one of the first attempts is due to Einsenhart~\cite{Eisenhart}, who established that trajectories of conservative systems have a correspondence with geodesics in a Riemannian manifold.  Following this line, Duval et al showed how to obtain a geometric structure for Newtonian physics, through Bargmann's structures~\cite{Duval1}.  This led to the Newton-Cartan theory which, when analyzed in the flat space, gives rise to a covariant formulation for the Schrödinger equation.  On the other hand, it is well known that the Poincaré group gives rise to the Galilei group through the well-known process  of Inönü-Wigner contraction\cite{Inonu}.  Thus Lévy-Leblond classified irreducible unitary representations of the Galilei group~\cite{levy}, in addition together with Le Bellac to obtaining two non-relativistic limits for electromagnetism, this made it possible to argue that spin has a Galilean nature \cite{levy1}.  Another important contribution by him was to obtain the non-relativistic Dirac equation~\cite{levy2}. In 1967, Pinski had constructed a similar tensor formulation based in the Galilean group~\cite{pinski}, but was only with Takahashi that a systematic theory using Lie algebras was developed. Takahashi et al presented a covariant formulation for the Galilei group based on the direct representations of that group~\cite{takahashi1, takahashi2, omote}, in contrast to the formulation based on the Bargmann structures that establishes the Galilei group from Poincaré group.  Both the formulation of Takahashi et al and that of Duval et al retrieve the results of Lévy-Leblond~\cite{Duval2}, but we believe that the Galilean covariancy obtained directly from the study of the representations of the Galilei group is more powerful. Takahashi's formulation is very different from those developed in the wake of Kaluza-Klein's ideas, despite the fact that this one has 5 dimensions. Here we will use an analogous approach to Takahashi Galilean covariance~\cite{takahashi1,omote, santana} instead of Duval's~\cite{duval2014}. There is an interesting duality between these two limits and this may be explored using the Galilean covariance formalism.

This paper is organized as follows, in section \ref{Carroll Group} the Carroll space embedded in the (4,1) de Sitter space is constructed. It is also shown that this space has an associated group that is an extended Carroll group. In section \ref{rep} the quantum representations of the extended Carroll group, namely the scalar and spinorial representations of Carroll fields are presented. In section \ref{max}  the Carrollian electric and magnetic limits using two different embedding of de Sitter space is showed. Finally, the conclusions are given in section \ref{conc}.

\section{The Carroll Group}\label{Carroll Group}
The five-dimensional manifold with the metric 
\begin{equation}
   g_{\mu\nu} = \left(\begin{array}{ccccc}
   1&0&0&0&0\\
   0&1&0&0&0\\
   0&0&1&0&0\\
   0&0&0&0&-1\\
   0&0&0&-1&0
   \end{array}\right),
\end{equation}
is a (4,1) de Sitter space under the transformation $g_{\mu\nu}=U_\mu^{\text{ \,}\alpha}\,\eta_{\alpha\beta} U^\beta_{\text{ \;}\nu}$, where $(\eta_{\alpha\beta})=(1,1,1,-1,1)$. 
This is easily seen by choosing the representation of $U^\mu_\nu$ as
\begin{equation}
  U^\mu_{\text{ \;}\nu} = \left(\begin{array}{ccccc}
   1&0&0&0&0\\
   0&1&0&0&0\\
   0&0&1&0&0\\
   0&0&0&\frac{1}{\sqrt{2}}&\frac{1}{\sqrt{2}}\\
   0&0&0&\frac{1}{\sqrt{2}}&-\frac{1}{\sqrt{2}}
   \end{array}\right).
\end{equation}
The associated group of this manifold has a Lie algebra defined by the following commutation rule,
\begin{equation}
\begin{split}
&\left[M_{\mu\nu},M_{\rho\sigma}\right]=-i(g_{\nu\rho}M_{\mu\sigma}-g_{\mu\rho}M_{\nu\sigma}+g_{\mu\sigma}M_{\nu\rho}-g_{\mu\sigma}M_{\nu\rho}),\\
\\
&\left[P_\mu, M_{\rho\sigma}\right]=-i(g_{\mu\rho}P_\sigma-g_{\mu\sigma}P_\rho),\\
\\
&\left[P_\mu, P_{\sigma}\right]=0.
\end{split}
\end{equation}
We can rewrite the generators, in a decomposition of (3+1+1) dimensions, as
\begin{eqnarray}
J_i&=&\frac{1}{2}\epsilon_{ijk}M_{jk},\nonumber\\
K_i&=&M_{5i},\nonumber\\
C_i&=&M_{4i},\nonumber\\
D&=&M_{54}.\label{rel_com}
\end{eqnarray}
Thus, the commutation relation becomes,
\begin{eqnarray}\label{nnc}
\begin{aligned}[c]
\left[J_i,J_j\right]&=i\epsilon_{ijk}J_k,\\
\left[J_i,C_j\right]&=i\epsilon_{ijk}C_k,\\
\left[D,K_i\right]&=iK_i,\\
\left[P_4,D\right]&=iP_4,\\
\left[P_i,K_j\right]&=i\delta_{ij}P_5,\\
\left[P_4,K_i\right]&=iP_i,\\
\left[D,P_5\right]&=iP_5,
\end{aligned}
\qquad\qquad
\begin{aligned}[c]
\left[J_i,K_j\right]&=i\epsilon_{ijk}K_k,\\
\left[K_i,C_j\right]&=i\delta_{ij}D+i\epsilon_{ijk}J_k,\\
\left[C_i,D\right]&=iC_i,\\
\left[J_i,P_j\right]&=i\epsilon_{ijk}P_k,\\
\left[P_i,C_j\right]&=i\delta_{ij}P_4,\\
\left[P_5,C_i\right]&=iP_i.\\
\end{aligned}
\end{eqnarray}
It is known that the Lie algebra of the extended Galilei group in $\mathcal{R}^3\times \mathcal{R}$ is a sub-agebra of this algebra, with $J_i$, as generators of rotations, $K_i$ of the pure Galilei boosts, and $P_\mu$ spacial and temporal translations, being $P_5$, in this context, a Casimir invariant associate with the mass, $P_5=-mI$, where $I$ is the identity matrix~\cite{takahashi1,omote,santana}. Another sub-algebra follows by setting the only non-zero commutation relations as    
\begin{eqnarray}\label{nnc_2}
\begin{aligned}[c]
\left[J_i,J_j\right]&=i\epsilon_{ijk}J_k,\\
\left[J_i,C_j\right]&=i\epsilon_{ijk}C_k,\\
\left[J_i,P_j\right]&=i\epsilon_{ijk}P_k,\\
\end{aligned}
\qquad\qquad
\begin{aligned}[c]
\left[P_i,C_j\right]&=i\delta_{ij}P_4,\\
\left[P_5,C_i\right]&=iP_i.\\
\end{aligned}
\end{eqnarray}
This is the algebra of Carroll group $\mathcal{C}$ with the addition of  $\left[P_5,C_i\right]=iP_i$, that comes naturally of the structure of the five-dimensional manifold. In this context $P_5$ is not a Casimir invariant as is in the case of Galilei group $\mathcal{G}$. Indeed the Casimir invariants of this algebra are
\begin{subequations}
\begin{eqnarray}
I_1&=&p^\mu p_\mu\label{I-1},\\
I_2&=&p_4,\label{I-2}\\
I_3&=&W_{4\mu}W^4_\mu,\label{I-3}
\end{eqnarray}
\end{subequations}
where $W^{\mu\nu}$ is the 5-dimensional Pauli-Lubanski matrix.

The transformations associated with this algebra are
\begin{subequations}
\begin{eqnarray}
      \overline{q}^i &=& R^i_jq^j-\frac{v^i}{c^\prime}q^5+ a^i,\label{C-1}\\
       \overline{q}^4&=& q^4-(R^i_jq^j)\frac{v_i}{{c^\prime}}+\frac{1}{2}\frac{\textbf{v}^{2}}{{c^\prime}^2}q^5+a^4,\label{C-2}\\
       \overline{q}^5 &=& q^5+a^5\label{C-3},
\end{eqnarray}
and
\begin{eqnarray}\label{emc}
	\overline{p}^i&=&R^i_{\text{\;}j}p^j-\frac{v^i}{c'}p^5,\\
	\overline{p}^4&=&p^4-\frac{v_i}{{c'}}(R^i_jp^j)
	+\frac{1}{2}\frac{\textbf{v}^{2}}{{c'}^2}p^5,\\
	\overline{p}^5&=&p^5,
\end{eqnarray}
\end{subequations}
where $c^\prime$ is a constant of velocity.

Choosing $q^\mu=(\textbf{q},c' t,s)$ and $p^\mu=(\textbf{p}, m\alpha c', \frac{E}{c'})$, where $s\equiv \frac{\textbf{q}^2}{2c^\prime t}$, these are the Carroll transformations in five dimensions (from now on we treat $c^\prime=1$).

It is worthy noting that even though $p_5$ can not be interpreted as the invariant mass, it, nevertheless, carries mass information.
\section{Representation of Quantum Mechanics}\label{rep}
In this section we will construct the representations of quantum mechanics of the extended Carroll group.
\subsection*{Scalar Representation}
For the scalar representation we take the invariants  $I_1$ \eqref{I-1} and $I_2$ \eqref{I-2} and apply to a function $\psi$, and using the correspondence relation $p^\mu=i\partial^\mu$ we have
\begin{eqnarray}\label{KG-1}
    \left\{\begin{array}{ll}
     \partial_{\mu}\partial^{\mu}\Psi=k^2\Psi\\
      \partial_{4}\Psi=-iE\Psi
       \end{array}\right.,
\end{eqnarray}
where $k$ and $E$ are constants. This is a non relativistic Klein-Gordon-like equation with carrollian symmetry.

 Using $\Psi(x^{\mu})=\exp\Big((-i(E t+m\,\alpha\;s)) \psi(\textbf{x})\Big)$, we have
\begin{equation}\label{DKP-25}
\begin{aligned}
        -\nabla^{2}\psi(\textbf{x})=2m\alpha E\psi(\textbf{x})
    \end{aligned}
\end{equation}
In this context the 5-current is
\begin{eqnarray}
j^\mu(x)=-i\Big(\psi^*(x)\partial^\mu\psi(x)-\partial^\mu(\psi^*(x))\psi(x)\Big),
\end{eqnarray}
and is conserved because the 5-divergence is null, ie
\begin{eqnarray}
\partial_\mu j^\mu=0.\label{current}
\end{eqnarray}
So the 5-current is equivalent to the usual 4-current,
\begin{eqnarray}
\textbf{j}(q)&=&-\frac{i}{2m\alpha}\Big[\Psi^*(q)\nabla\big(\Psi(q)\big)-\nabla\big(\Psi^*(q)\big)\Psi(q)\Big],\\
\nonumber\\
j^4=\rho(q)&=&-\frac{i}{2m\alpha}\Big[-\Psi^*(q)\partial_s\big(\Psi(q)\big)+\partial_s\big(\Psi^*(q)\big)\Psi(q)\Big]=|\Psi|^2,
\end{eqnarray}
where $ \textbf{j}(x) $ is the probability current and $ \rho (q) $ is the probability density.
\subsection*{Spinor Representation}
In this context, we present a construction of the spin 1/2 wave equation, defining a new quadrivector $\gamma^\mu$ such that,
\begin{eqnarray}
(\partial_{\mu}\partial^{\mu}-k^2)=(\gamma^{\mu}\partial_{\mu}+k)(\gamma^{\nu}\partial_{\nu}-k),\label{eq:dirac1}
\end{eqnarray}
for \eqref{eq:dirac1} to be valid $\gamma^\mu$ must obey the Clifford algebra, that is
\begin{eqnarray}
 \{\gamma^\mu,\gamma^\nu\}=2g^{\mu\nu},\label{eq:cliff}
\end{eqnarray}
where $ g^{\mu\nu} $ is our penta-dimensional metric.
Taking the plus-sign bracket and operating in the $ \Psi(x) $ wave function, we get
\begin{eqnarray}
(\gamma^{\mu}\partial_{\mu} + k) \Psi(x) = 0. \label{eq: dirac2}
\end{eqnarray}
For convenience, we will use the following representations of $ \gamma^\mu$
\begin{eqnarray*}
\gamma^i=
\left(\begin{array}{cc}
\sigma^i&0\\
0&-\sigma^i
\end{array}\right),\qquad
\gamma^4=
\left(\begin{array}{cc}
0&0\\
-\sqrt{2}&0
\end{array}\right),\qquad
\gamma^5=
\left(\begin{array}{cc}
0&\sqrt{2}\\
0&0
\end{array}\right).
\end{eqnarray*}
where $\sigma^i$ are Pauli's arrays and $\sqrt{2}$ is the 2x2 identity matrix multiplied by $\sqrt{2}$. We can write the $\Psi$ object, as
\begin{eqnarray*}
\Psi=\left(\begin{array}{cc}
\varphi(\textbf{x},x^4,x^5)\\
\chi(\textbf{x},x^4,x^5)
\end{array}\right),
\end{eqnarray*}
where $ \varphi $ and $ \chi $ are 2-spinors dependent on $ x^\mu; \mu = 1, ..., 5 $. Therefore, in the representation where $k = 0$, Eq.~\eqref{eq: dirac2} is reduced to
\begin{subequations}\label{eq:Levy-Leblond}
\begin{eqnarray}
\boldsymbol{\sigma}\cdot\nabla\varphi+\sqrt{2}\partial_s\chi=0,\label{eq:Levy-Leblond1}\\
\nonumber\\
\sqrt{2}\partial_t\varphi+\boldsymbol{\sigma}\cdot\nabla\chi=0.\label{eq:Levy-Leblond2}
\end{eqnarray}
\end{subequations}
Eq.~\eqref{eq:Levy-Leblond} are the Carroll-L\'evy-Leblond equations.
The 5-current is
\begin{eqnarray}
j^\mu(x)=\frac{1}{\sqrt{2}i}\left[\overline{\psi}(x)\gamma^\mu\psi(x)\right],
\end{eqnarray}
where $\overline{\psi}=\psi^\dagger\zeta$, with
\begin{equation*}
\zeta=-\frac{i}{\sqrt{2}}(\gamma^4+\gamma^5)=
\left(\begin{array}{cc}
0&-i\\
i&0
\end{array}\right),
\end{equation*}
and $j^\mu$ is conserved, the 5-divergence is null
\begin{eqnarray}
\partial_{\mu}J^\mu=0.\nonumber
\end{eqnarray}
In terms of $\varphi$ e $\chi$
\begin{eqnarray}
j^i&=&\frac{1}{\sqrt{2}}\left[\chi^\dagger\boldsymbol{\sigma}\varphi+\varphi^\dagger\boldsymbol{\sigma}\chi\right],\\
\nonumber\\
\nonumber\\
j^4&=&\varphi^\dagger\varphi,\qquad\qquad j^5=\chi^\dagger\chi,
\end{eqnarray}
using Eq.\eqref{KG-1}, \eqref{eq:Levy-Leblond1} e \eqref{eq:Levy-Leblond2} we have
\begin{eqnarray*}
j^i&=&-\frac{i}{2m\alpha}\partial_i\left[\varphi^\dagger(x)\partial^i\varphi(x)-\partial^i(\varphi^\dagger(x))\varphi(x)\right]+\frac{1}{2m\alpha}\partial^j\left[\varphi^\dagger\sigma^k\varphi\right]\epsilon_{ijk},\\
\text{and}\\
\partial_5j^5&=&\partial_s(\chi^\dagger\chi)=0.
\end{eqnarray*}
The first term in $j^i$ represents the probability current, given by Eq. \eqref{current}, and the second is associated with the spin current, which results in the correct intrinsic magnetic moment value of the particle.

\section{The Electric and Magnetic Limits}\label{max}
In this section we show the electric and magnetic Carrollian limits of Maxwell equations \cite{duval2014}, using specific immersions.

In terms of the Faraday tensor, $F_{AB}$, the Maxwell equations are

\begin{equation}
\partial^AF_{AB}=j_B, \label{eqf1}
\end{equation}
\begin{equation}
\partial_MF_{AB}+\partial_AF_{BM}+\partial_BF_{MA}=0.\label{eqf2}
\end{equation}

To obtain the differential equation in terms of the Electric and Magnetic fields we use the explicit form of the five-dimensional Faraday tensor
\begin{eqnarray}
F_{AB}=
\left(\begin{array}{ccccc}
 0&B_2&-B_2&c_1&d_1\\
   -B_3&0&B_1&c_2&d_2\\
 B_2&-B_1&0&-c_3&d_2\\
   -c_1&-c_2&-c_3&0&R\\
    -d_1&-d_2&-d_3&-R&0
   \end{array}\right).
\end{eqnarray} 
which applied to equations Eq. \eqref{eqf1} and \eqref{eqf2} results in
\begin{eqnarray}
\begin{aligned}[c]
&\nabla\cdot \textbf{B}=0,\\
&\nabla\times \textbf{c}+\partial_4\textbf{B}=0,\\
&\nabla\times \textbf{d} + \partial_5\textbf{B}=0,
\end{aligned}
\qquad
\begin{aligned}[c]
&\nabla\cdot \textbf{c}=j_4+\partial_4R,\\
&\nabla\cdot \textbf{d}=j_5+\partial_5R,\\
&\nabla\times\textbf{B}-\partial_4\textbf{d}-\partial_5\textbf{c}=\textbf{j}.
\end{aligned}
\end{eqnarray}
The fields are given by
\begin{eqnarray}
\begin{aligned}[c]
\textbf{c}=\nabla A_4-\partial_4\textbf{A},\\
\textbf{d}=\nabla A_5-\partial_5\textbf{A},
\end{aligned}
\qquad
\begin{aligned}[c]
\textbf{B}=\nabla\times\textbf{A}
\end{aligned}
\end{eqnarray}
where $\textbf{A}$ is the vector potential. Letting $\textbf{c}=0$, $R=0$  and $\textbf{d}=\textbf{E}$, is the electric field.

We can obtain the Carrollian magnetic limit if we choose the following immersions
\begin{eqnarray*}
x^A = (\textbf{x},t,0),\qquad A^A=(\textbf{A},0,-\phi).
\end{eqnarray*}
Thus, under Carrollian boost, we have
$$
\bar{\textbf{x}}=\textbf{x},\qquad \bar{t}= t-\textbf{v}\cdot\textbf{x},\qquad \bar{x}^5=0,
$$
So, $\partial_4=\partial_t$ and, as $A^A$ is a massless particle it is independent of $\alpha$, then $\partial_5\textbf{A}=0$. The resulting Maxwell equations  are
\begin{eqnarray}
\begin{aligned}[c]
&\nabla\cdot \textbf{E}=\rho,\\
&\nabla\times \textbf{E}=0,\\
\end{aligned}
\qquad
\begin{aligned}[c]
&\nabla\cdot \textbf{B}=0,\\
&\nabla\times\textbf{B}-\partial_t\textbf{E}=\textbf{j}.
\end{aligned}
\end{eqnarray}
\begin{eqnarray*}
\textbf{E}=-\nabla\phi,\qquad\textbf{B}=\nabla\times\textbf{A}.
\end{eqnarray*}
The choice of gauge $\partial_\mu A^\mu = 0$ reduces to $\nabla\cdot\textbf{A} = -\partial_t\phi$, the Lorenz gauge. The movement of electric charges is capable of creating a magnetic field, but a time-varying magnetic field would not create an electric field.

In te case of the Carrollian electric limit we made choose the following immersions
\begin{eqnarray*}
x^A = (\textbf{x},0,t),\qquad A^A=(\textbf{A},0,-\phi).
\end{eqnarray*}
Thus, under Carrollian boost, we have
$$
\bar{\textbf{x}}=\textbf{x}+\textbf{v}t,\qquad \bar{x}^4=0,\qquad \bar{t}= t,
$$ 
and the obtained Maxwell equations are
\begin{eqnarray}
\begin{aligned}[c]
&\nabla\cdot \textbf{E}=\rho,\\
&\nabla\times \textbf{E}+\partial_t\textbf{B}=0,\\
\end{aligned}
\qquad
\begin{aligned}[c]
&\nabla\cdot \textbf{B}=0,\\
&\nabla\times\textbf{B}=\textbf{j}.
\end{aligned}
\end{eqnarray}
\begin{eqnarray*}
\textbf{E}=-\nabla\phi-\partial_t\textbf{A},\qquad\textbf{B}=\nabla\times\textbf{A}.
\end{eqnarray*}
The choice of gauge $\partial_\mu A^\mu = 0$ reduces to $\nabla\cdot\textbf{A} = 0$, the Coulomb gauge. The temporal variation of the magnetic field creates an electric field but not the other way around. This results shown here are in accordance with the literature~\cite{duval2014}. We note that the Carrollian electric limit has Galilean symmetry in its coordinates, as the Galilean magnetic limit has Carrollian symmetry~\cite{de1999poincare}.

The transformation of the fields are
\begin{equation*}
    \begin{aligned}
 &\bar{\textbf{E}}=\textbf{E}.\\
 &\bar{\textbf{B}}=\textbf{B}-\textbf{v}\times \textbf{E}.
    \end{aligned}\;\Bigg\}
    \rightarrow\text{electric limit,}\quad
    \begin{aligned}
  &   \bar{\textbf{E}}=\textbf{E}+\textbf{v}\times \textbf{B}.\\
 &\bar{\textbf{B}}=\textbf{B}.
    \end{aligned}\;\Bigg\}\rightarrow\text{magnetic limit.}
\end{equation*}
\section{The interpretation of $\alpha$}\label{alpha}
Letting $\alpha=1$ we have for the mass shell condition
\begin{equation}
    p^2-2mE=0,
\end{equation}
this has the same form of Galilean mass shell condition. The difference over these two symmetries are that in the case of Carroll as the energy is invariant $\alpha$ varies with the relative velocity. In this way, Carroll particles can indeed move. Thus even though the momentum, $p$, transforms like \eqref{emc}, the energy,  $\frac{p^2}{2m}$, is invariant. If, in a inertial frame, the momentum is zero, we have the special case~\cite{duval2014,duval2017}
\begin{equation*}
    E^2-m^2=0,
\end{equation*}
where we have reintroduced the rest energy. Setting the speed of light $c=1$, thus instead of the limit $c\rightarrow0$, we will have $v\gg 1$, thus a Carrollian particle in this context will describe a tachyon. Here, the $\alpha$ parameter can be interpreted as a drag and the tachyon will acquire Carrollian symmetry in this limit. A experiment in the context of dual gravity/fluid can be proposed to study this drag, in this context a soliton should acquire Carrolian-like symmetry when $v\gg c'$, where $c'$ is the sound speed in the fluid.
\section{conclusion}\label{conc}

In this article we describe Carroll's transformations in a covariant way, analogously to what we have for Galilei's transformations.  The Carroll transformation is a non-Galilean limit on Lorentz's transformations, so it is natural to propose its generalization in 5 dimensions. We apply this description to the scalar field, the 1/2 spin field and electromagnetism.  We interpret the alpha parameter which has a close relationship with the Corrollian particle, this parameter is dependent on the reference system and indicates a tachyonic particle.

\section*{Aknowlegdments}

This work is supported by CAPES and CNPq of Brazil. 

I also would like to thank Professors J. M. Levy-Leblond and J. P. Gazeau for the useful correspondences.

\end{document}